\documentclass[reprint,superscriptaddress,twocolumn,aps,pre,showpacs]{revtex4-1}
\usepackage{amsbsy,amssymb,amsmath,bm,ulem}
\usepackage{graphicx}

\begin{document}
\title{Modelling nonlocal electrodynamics in superconducting films: The case of a concave corner}

\author{J. I. Vestg{\aa}rden}
\affiliation{Department of Physics, University of Oslo, P. O. Box
1048 Blindern, 0316 Oslo, Norway}
\author{T. H. Johansen}
\affiliation{Department of Physics, University of Oslo, P. O. Box
1048 Blindern, 0316 Oslo, Norway}
\affiliation{Institute for Superconducting and Electronic Materials,
University of Wollongong, Northfields Avenue, Wollongong,
NSW 2522, Australia}
\affiliation{Centre for Advanced Study at The Norwegian Academy
of Science and Letters, Drammensveien 78, 0271 Oslo, Norway}

\begin{abstract}
We consider magnetic flux penetration in a superconducting film with a
concave corner. Unlike convex corners, where the current flow pattern
is easily constructed from Bean's critical state model, the current
flow pattern at a concave corner is highly nontrivial.  To address the
problem, we do a numerical flux creep simulation, where particular
attention is paid to efficient handling of the non-local
electrodynamics, characteristic of superconducting films in the transverse
geometry.  We find that the current stream lines at the concave corner
are close to circular, but the small deviation from exact circles
ensure that the electric field is finite and continuous.  Yet, the
electric field is, as expected, very high at the concave corner.  At
low fields, the critical state penetration is deeper 
from the concave corner than from the straight edges, which is a
consequence of the electrodynamic
non-locality.  A magneto-optical experiment on a YBa$_2$Cu$_3$O$_x$
displays an almost perfect match with the magnetic flux distribution
from the simulation, hence verifying the necessity of including
electrodynamic non-locality in the modelling of superconducting
thin films.
\end{abstract}



\maketitle 

\section{Introduction}
The macroscopic magnetic properties of type-II superconductors can to
a large extent be described by the critical state model, first
formulated by Bean \cite{bean62,bean64}. It is based on the assumption
that the magnitude of the current cannot exceed the critical current
density $j_c$, thus limiting the ability of the superconductor to
carry transport current or to shield applied magnetic fields.  
The original critical state model was
formulated for bulk samples, but it has later been extended to thin
films, where additional shielding currents $j<j_c$ flow in the region
beyond the flux penetration front. Analytical results have been found
for an infinite strip \cite{norris69,brandt93} and circular disk
\cite{mikheenko93,clem94}, while for less symmetric shapes or
disconnected geometries the critical state has only been determined
numerically \cite{prigozhin98}.

Dynamics is disregarded in Bean's critical state model, in the sense
that the response to an applied magnetic field is instantaneous. In
particular, the model does not take into account flux creep, a process
acting in any real type-II superconductor, and is particularly
pronounced in high-$T_c$ superconductors.  To model flux creep, time
must be taken explicitly into consideration. Conventionally, this is
done by assuming a highly nonlinear $E-J$ relation before solving
Maxwell's equations. For thin films, the main obstacle for an
efficient implementation of a numerical scheme, is the hand\-ling of
the boundary conditions, given the nonlocality of the governing
equations. Brandt has derived solutions for selected geometries, such
as rectangles \cite{brandt95}, disks and rings \cite{brandt97}, or
arbitrary shape \cite{brandt01}, based on a matrix inversion
method. Although being accurate, these solutions scale poorly with
system size and need $O(N^2)$ operation for each time step, where $N$
is the number of discrete points in the spatial grid.  Better scaling
properties can be obtained by the conjugate gradient method, which
scales like $O(N^{1.4})$ \cite{wijngaarden98,loerincz04} or a hybrid
real space - Fourier space method scaling as $O(N\log(N))$
\cite{vestgarden11,vestgarden12}.

One remarkable property of the Bean model is that the current stream
line pattern can be drawn simply by adding lines with constant
spacing, starting from the contour of the sample edge. When there is a
constriction, the stream lines must adapt by bending, which leads to
formation of so called d-lines, i.e., lines in the flow pattern where
the stream lines change direction discontinuously. These d-lines are
recognized also in the corresponding magnetic flux distribution, due
to an almost total local suppression of the magnetic field in their
vicinity.  By examination of measured flux distributions, e.g.,
obtained by magneto-optical imaging (MOI), it has been found that the
Bean model to a large extent explains the flux penetration patterns of
hard superconductors.  For example, the d-line created by a circular
nonconducting hole (antidot) is parabolic
\cite{CCS,schuster94,vestgarden08}, a row of antidots give
asymptotically straight d-lines at an angle given by the antidot
fraction \cite{vestgarden12}, while a convex right-angled corner,
e.g., as in a rectangular sample, gives straight d-lines making
45$^\circ$ with the meeting edges \cite{schuster95}.

\begin{figure}[t]
  \centering
  \includegraphics[width=8.5cm]{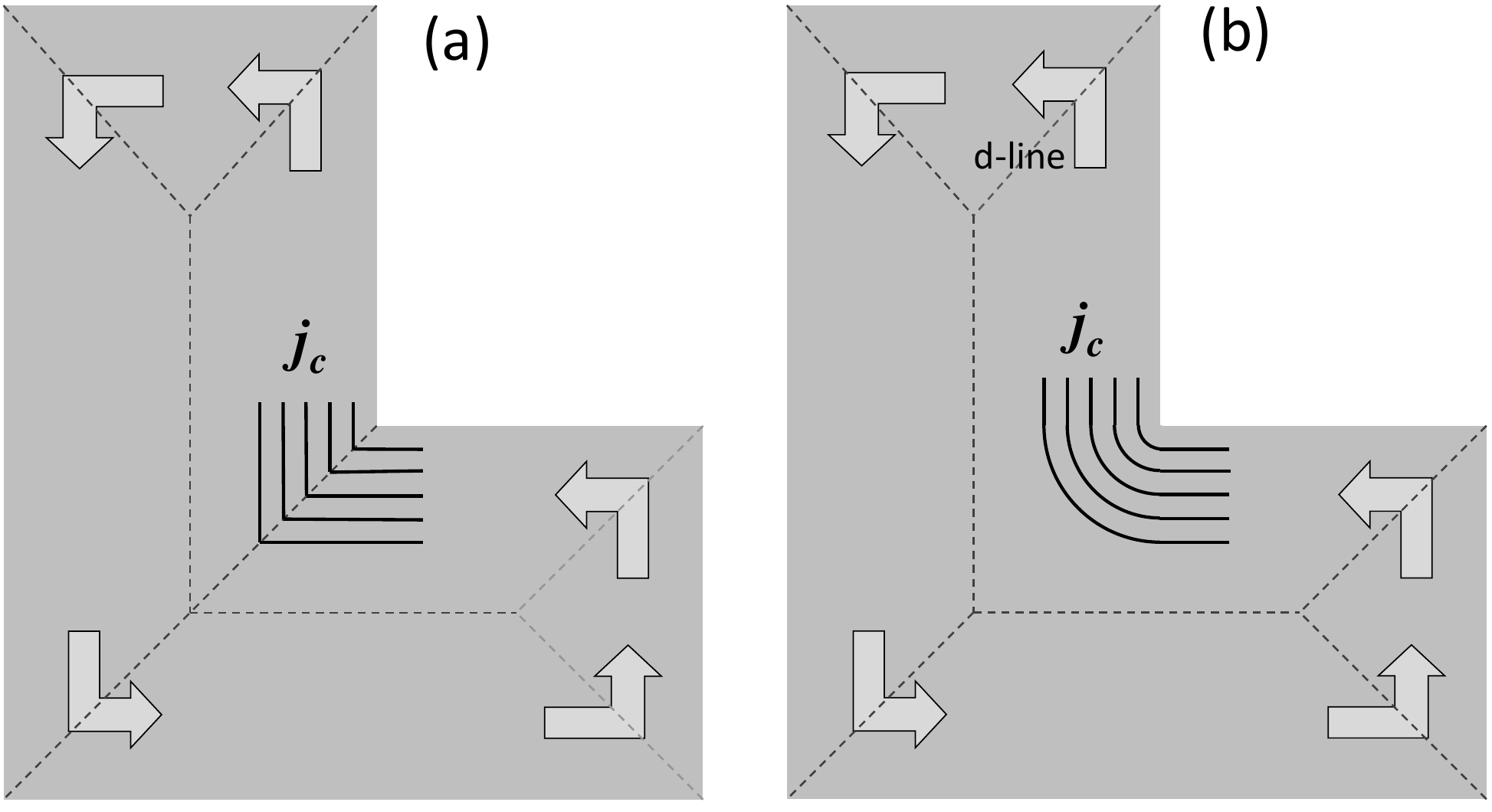} \
  \caption{
    \label{fig:bean}
 Two different stream line patterns near a concave corner with flow of a 
constant-magnitude  
current density, $j_c$, as in the Bean model. 
  }
\end{figure}

Despite its vast success, the Bean model procedure for drawing
critical current stream-line patterns is ambiguous at a concave
corner.  This is apparent in Fig.~\ref{fig:bean}, demonstrating two
possible ways to join stream lines as they pass a 90$^\circ$ concave
corner. In (a) the equidistant stream lines are straight and meet at a
$45^\circ$ d-line, as in the case of a convex corner. In (b) the
current stream lines consist partly of circular segments, and there
are no traces of d-lines.  The constructions (a) and (b) are just two
out of infinitely many possible solutions that conserve current and
give constant current density \cite{friesen01}.  However, only one
construction can be correct, and to identify it a more comprehensive
analysis must be made.

Schuster, Kuhn and Brandt \cite{schuster96} investigated the flux and
current distributions in a cross-shaped superconductor, a geometry
that contains both convex and concave corners. Their MOI experiment
showed no d-lines at the concave corners, hence excluding the sharp
turn illustrated in Fig.~\ref{fig:bean}a. An accompanying flux creep
simulation gave stream lines close to circular, as in
Fig.~\ref{fig:bean}b.  At the same time, a theoretical estimate showed
that exactly circular stream lines diverge as $E\propto 1/r$ close to
the corner and the $E$-field will make a jump where the circular
stream lines connect with the straight lines running parallel to the
edges. As pointed out by Ref. \cite{gurevich00} this is an indication
that the stream line pattern is unphysical.  Thus, the stream lines at
a concave corner cannot be perfectly circular.

Gurevich and Friesen \cite{gurevich00,friesen01} calculated
analytically the stream lines in the steady state using the powerful
and general hodograph method.  Their solution showed that, in the
limit of negligible flux creep, neither Fig.\ref{fig:bean} (a) nor (b)
are correctly representing the stream line pattern near a concave
corner. However, their solution did not include the history of the
magnetization process leading to the current distribution.  Neither
did it include the specific non-locality effects crucially important
in the transverse geometry. Thus, it still remains an open question
what is the correct current stream line pattern in a thin
superconductor with a concave corner, and subjected to a perpendicular
magnetic field.

This work considers in detail the flux penetration and current flow
in a thin superconductor having the simplest shape that includes
a concave corner, namely that shown in Fig.~\ref{fig:bean}.  A general
simulation method for flux dynamics in thin samples of almost any shape
is described in Sec. II. The approach takes special care to
model the non-locality of the equations in an efficient way.  Section III
reports and discusses the simulations results 
which include distributions of magnetic field, shielding current and
electrical field. A direct comparison with MOI experiments on a 
YBa$_2$Cu$_3$O$_x$ film in an increasing
applied magnetic field is reported in Sec. IV. 
 Finally, Sec.~V gives the conclusions.

\section{Simulation Model}

The numerical scheme is quite general and can be applied to thin planar superconductors
of arbitrary shape, provided the thickness $d$ is much smaller than
any lateral dimension, and the whole sample can be embedded inside a
rectangular area \cite{vestgarden12}. We also require that the 
external field $H_a$ is applied in the $z$-direction, transverse to the sample plane. 
Then, the flux dynamics is found by solving Maxwell's equations in the quasi-static
(eddy current) approximation, where the superconducting
properties enters the equations through a highly nonlinear $E-J$ relations that
characterizes the sharp vortex depinning transition happening when the sheet
current $J$ approaches the value of the critical sheet current $J_c=dj_c$. A realistic
approximation for many materials is a power law \cite{brandt95}
\begin{equation}
  \mathbf E = \rho \mathbf J/d,\quad\rho=\rho_0 \left(J/J_c \right)^{n-1}
  ,
  \label{power-law-Ej}
\end{equation}
where $\mathbf E$ is the electric field, $\rho$ is 
the resistivity, and $\rho_0$ is a resistivity constant.
The creep exponent $n$ is usually large, with 
the Bean model corresponding to the limit $n\to \infty$.
Note that the model also works for an ohmic conductor, where $n=1$.

Since current is conserved, $\nabla\cdot \mathbf J = 0$, 
we can introduce the local magnetization $g=g(\mathbf r,t)$ as
\begin{equation}
  \label{defg}
  \frac{\partial g}{\partial y}=J_x
  ,
  ~~~  
  \frac{\partial g}{\partial x}=-J_y
  ,
\end{equation}
where $\mathbf r = (x,y)$.  Outside the sample, $g \equiv 0$. 
The integral of $g$ gives the magnetic moment, $m=\int d^2rg(\mathbf r)$.

For quasistatic situation, the $B-J$ relation is given by the non-local 
Biot-Savart law, which can be rewritten to a $B_z-g$ relation as  
\begin{equation}
  \label{b-s}
  B_z/\mu_0=H_a + \hat Qg
  ,
\end{equation}
with the operator $\hat Q$ given by
\begin{equation}
  \label{hatQ}
  \hat Qg(\mathbf r) 
  = {\mathcal F}^{-1}\left[\frac{k}{2}\mathcal F\left[g(\mathbf r)\right]\right]
  ,
\end{equation}
where  $\mathcal F$ is the 2D spatial Fourier transform, $k=|\mathbf k|$,
and $\mathbf k$ is the wave-vector. The inverse relation is 
\begin{equation}
  \label{hatinvQ}
  \hat Q^{-1}\varphi(\mathbf r) 
  = {\mathcal F}^{-1}\left[\frac{2}{k}\mathcal F\left[\varphi(\mathbf r)\right]\right]
  ,
\end{equation}
where $\varphi$ is an auxiliary function. Both $\hat Q$ and $\hat Q^{-1}$ are direct product 
in Fourier space and can thus be calculated effectively.

By taking the time derivative of Eq.~\eqref{b-s} and inverting it, we get
\begin{equation}
  \label{dotg}
  \dot g = \hat Q^{-1}\left[\dot B_z/\mu_0- \dot H_a\right]
  .
\end{equation}
This equation is solved by discrete integration forward in time.
In order to carry out the time integration,
$\dot B_z$ must be known in the whole plane $z=0$ at time $t$
and, in this work, completely different approaches will be used to find
$\dot B_z$ within the superconductor and in the surrounding vacuum.

Starting with the superconductor itself,
it obeys the material law,  
Eq.~\eqref{power-law-Ej}, which when combined with
Faraday's law, $\dot B_z=- (\nabla \times \mathbf{E})_z$, 
gives
\begin{equation}
  \label{se2}
  \dot B_z = \nabla \cdot (\rho\nabla g)/d\, .
\end{equation}
From $g(\mathbf r,t)$ the gradient is readily calculated, and since
the result allows finding $\mathbf J(\mathbf r,t)$ from
Eq.~\eqref{defg}, also $\rho(\mathbf r,t)$ is determined from
Eq.~\eqref{power-law-Ej}. The task then is to find 
$\dot{B}_z$ in the non-superconducting parts, so that 
$\dot g=0$ outside the superconductor. This cannot be calculated efficiently using direct 
methods due to the nonlocal relation between $\dot{B}_z$ and $\dot{g}$.  
Instead we use an iterative procedure.

For all iteration steps, $i=1...s$, $\dot{B}_z^{(i)}$ is fixed inside the
superconductor by Eq.~\eqref{se2}. At $i=1$, an initial guess is made for $\dot{B}_z^{(i)}$ 
outside the sample, and $\dot g^{(i)}$ is
calculated from Eq.~\eqref{dotg}.  In general, this $\dot{g}^{(i)}$ does
not vanish outside the superconductor, but an improvement can be obtained by
\begin{equation}
  \label{iterative-Bz}
  \dot{B}_z^{(i+1)} = \dot{B}_z^{(i)}  -\mu_0\hat{Q}\hat{O} \dot{g}^{(i)} +C^{(i)} 
  .
\end{equation}
The projection operator $\hat O$ is unity outside the
superconductor and zero in the vacuum. Also, the output of the operation should 
be shifted to satisfy $\int d^2r\hat O\dot g^{(i)}=0$.
The constant $C^{(i)}$ is
determined by requiring flux conservation,
\begin{equation}
  \label{defC1}
  \int d^2r\, [\dot B_z^{(i+1)}-\mu_0 \dot H_a]=0  
  .
\end{equation}

Thus, at each iteration $(i)$, $\dot B_z^{(i+1)}$ is calculated for
the outside area.  The procedure is repeated until after $i=s$
iterations $\dot g^{(s)}$ becomes sufficiently uniform outside the
sample.  Then, $\dot g^{(s)}$ is inserted in Eq.~\eqref{dotg}, which
brings us to the next time step, where the whole iterative procedure
starts anew.

The non-dimensional form of the equations are particularly 
simple when the applied field is ramped with constant rate $\dot H_a\neq 0$
and $J_c$ and $n$ are both constants. We define the sheet current constant 
\begin{equation}
  J_0\equiv J_c\left(\frac{dw\mu_0\dot H_a}{\rho_0J_c}\right)^{1/n}
  ,
\end{equation}
where $w$ is some lateral length and the rest of the parameters 
have been defined previously. The time constant is defined as 
\begin{equation}
  t_0 \equiv J_0/\dot H_a
  .
\end{equation}
The dimensionless form of the material law, Eq.~\eqref{power-law-Ej}, becomes
\begin{equation}
  \label{nd-power-law-Ej}
  \tilde {\mathbf E} = \tilde\rho\tilde{\mathbf J},\quad \tilde \rho = \tilde J^{n-1}
  ,
\end{equation}
where $\tilde{\mathbf J} \equiv {\mathbf J}/J_0$ and $\tilde {\mathbf E} \equiv \mathbf Ed/(\rho_0 J_0)$.
The time evolution, Eq.~\eqref{dotg}, becomes
\begin{equation}
  \label{nd-dotg}  
  \frac{d\tilde g}{d\tilde t} = \tilde {\hat Q}\left[\frac{d\tilde B_z}{d\tilde t}-1\right]
  ,
\end{equation}
where $\tilde g \equiv g/(wJ_0)$, $\tilde {\hat Q} \equiv \hat Q/w$, $\tilde t \equiv t/t_0$, 
and $\tilde B_z \equiv B_z/(\mu_0J_0)$.
Finally, the Faraday law, Eq.~\eqref{se2}, becomes 
\begin{equation}
  \label{nd-se2}
  \frac{d\tilde B_z}{d\tilde t} = \tilde \nabla \cdot \left(\tilde \rho \tilde \nabla \tilde g\right)
  ,
\end{equation}
where $\tilde \nabla \equiv w\nabla$.  This means that the only free
parameter in the model is $n$.  A practical consequence of this is
that simulations only need to be run once for each geometry and each
value of $n$, and the result for any combination of dimensional
parameters can be found simply by rescaling the solution.

Several physical conclusions can be drawn directly from the
non-dimensional equations.  First, Eq.\eqref{nd-dotg} with
Eq.~\eqref{nd-se2} inserted is a nonlocal diffusion equation with $\tilde\rho$
as a very nonlinear diffusion constant. The non-locality prevents 
scaling solutions like in bulk \cite{shantsev02}, but will still give
plateaus in the current where $\tilde J \approx 1$.  Second, the level
of these plateaus in dimensional units is $J_0$, not $J_c$ as one
could naively expect.  Thus, $\tilde J>1$ does not imply $J>J_c$,
since $J_0<J_c$ for parameters corresponding to most type-II
superconductors.  Third, the current constant depends on ramp rate as
$J_0\propto \dot H_a^{1/n}$.  Such a ramp rate dependent current gives
also a flux penetration depending on ramp rate, and this has indeed
been measured by Ref.~\cite{koblischka97} in a  strip of high-$T_c$ superconductor.

In this paper the tildes are omitted when reporting the
results in dimensionless units.

We chose the lateral length scale $w$ as the half-width of the
shortest side of the superconducting corner.  The long sides of the
sample have length $6$. The sample is embedded in a $9\times 9$
square, which is discretized on a $512\times 512$ grid. Note that
since the geometry is very non-symmetric, only $25\%$ of the grip
points lay within the sample, the rest are in the vacuum, needed to
fulfil the boundary conditions.  Nevertheless, the method is fast and
the simulations of this paper can easily be run on a personal
computer.

\begin{figure}[t]
  \centering
  \includegraphics[width=8.5cm]{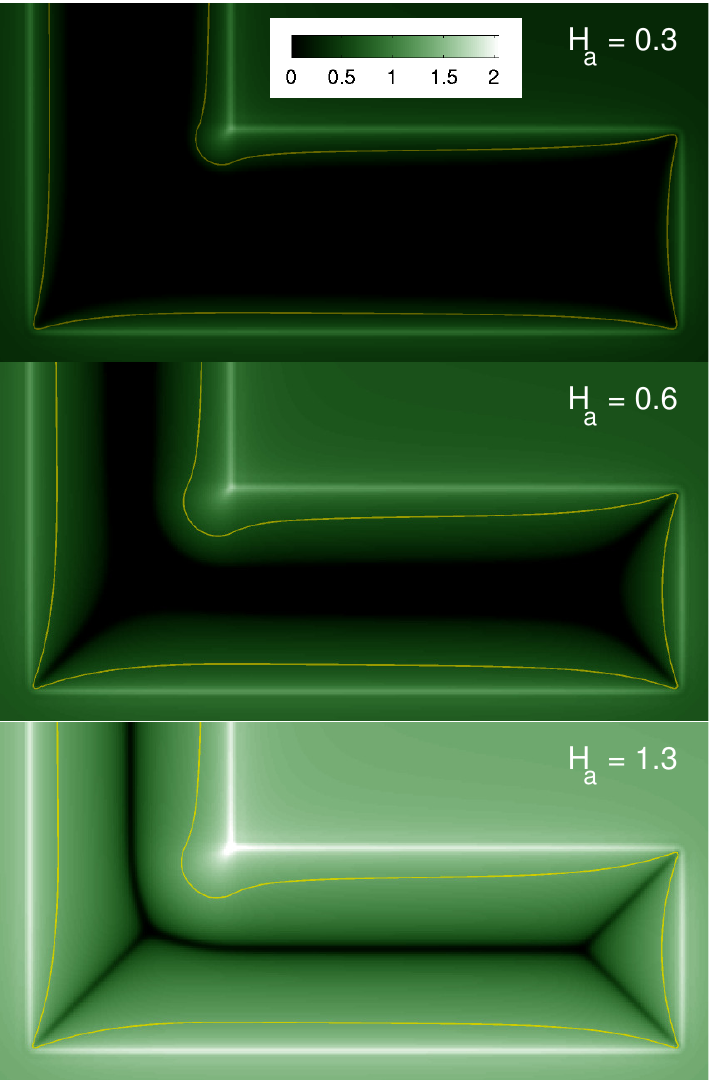} 
  \caption{
    \label{fig:H}
    Distribution of flux density $B_z$ at applied perpendicular fields $H_a=0.3$, 0.6 and 1.3.
    The images are color coded so that the brighter green the larger is $B_z$,
    see the color bar. Included in each panel is the contour line of a constant 
    flux density corresponding to the applied field,  $B_z = \mu_0 H_a$.
  }
\end{figure}

\begin{figure}[t]
  \centering  
  \includegraphics[width=8.5cm]{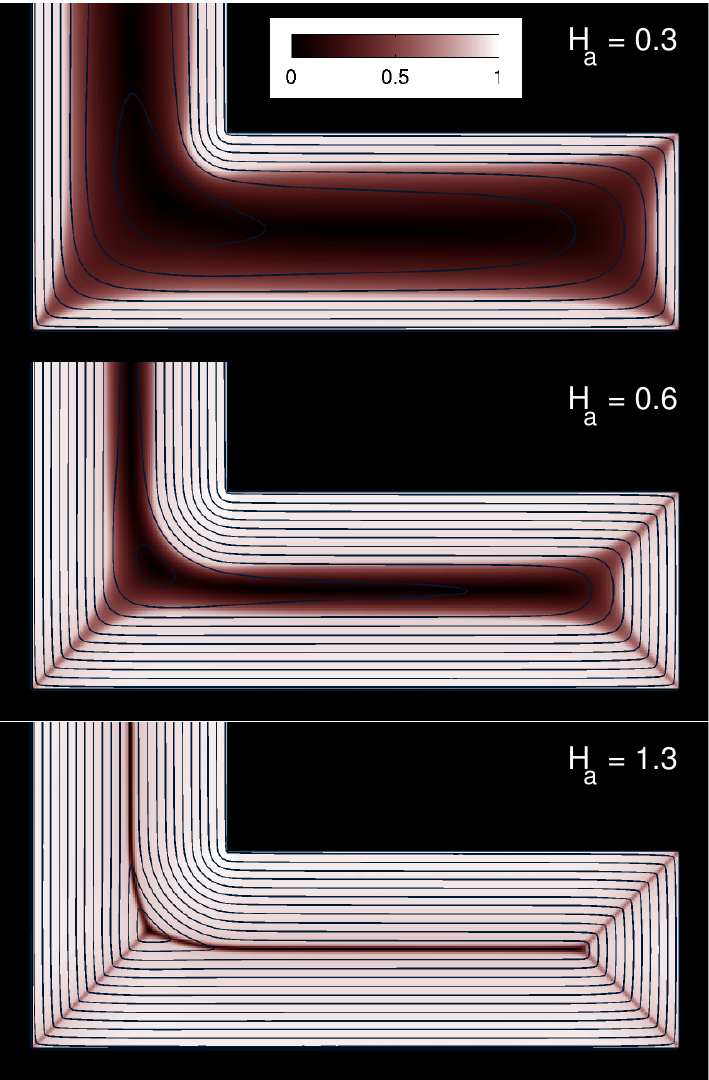} 
  \caption{
    \label{fig:j}
    Distribution of sheet current magnitude $J$ and current stream lines
    at applied perpendicular fields  $H_a=0.3$, 0.6 and 1.3.
    The images are color coded so that the brighter red the larger is $J$, 
    see the color bar.
  }
\end{figure}

\section{Simulation Results}
The initial state  is a flux free superconductor, and we use a
 flux creep exponent of $n=29$.
As the applied magnetic field is increased with constant rate
the magnetic flux gradually enters the sample from the edges,
as seen in Fig.~\ref{fig:H}, showing snap-shots of the
$B_z$-distribution at $H_a=0.3$, 0.6 and 1.3.
In the figure, the image brightness represents the 
magnitude of $B_z$, which everywhere is directed parallel to the applied field.

At $H_a=0.3$, the most visible feature is that the edges light up with
field values much larger than the applied field, while the sample
interior is black in an area representing the 
Meissner state where the flux is expelled from the superconductor.
Already at this low field, the concave corner shows enhanced field
values and deeper flux penetration as compared to the penetration from
the straight edges.

At $H_a=0.6$, the flux front has advanced considerably, and now the
concave corner is filled with a fan of enhanced flux density. In the
convex corners, the d-lines start developing, but they are not yet
distinct lines, but rather wedge-shaped regions of vanishing flux
density.

At $H_a=1.3$, the flux has penetrated essentially the entire sample,
and the superconductor is now described by a fully developed critical
state.  The d-lines are clearly seen in all convex corners, where they
make 45$^\circ$ angles with the meeting edges, as expected for a
superconductor with isotropic $J_c$. In the concave corner, on the
other hand, there is no sign of any d-line.  Quite the contrary, the
flux density is there higher than in the surrounding regions. In
particular, in the corner point itself the magnetic field is greatly
amplified.  Evidently, the current stream line pattern responsible for
this flux distribution must be close to Fig.~\ref{fig:bean}b, and
definitely not like in Fig.~\ref{fig:bean}a.

Figure~\ref{fig:j} shows color-coded images of the sheet current
magnitude, $J$, together with the stream lines of $\mathbf J$,
calculated as the contour lines of the local magnetization $g$.
At $H_a=0.3$ the critical state with $J\approx 1$ has only penetrated
a short distance from the edge, while most of the sample is in the
Meissner state, where $J<1$, as expected for a thin superconductor
in a small magnetic field.  The outermost stream lines can
be followed around the sample perimeter where they, along each straight
edge segment, stack with equal spacing, in accordance with the critical
state model.  At the concave corner the stream lines turn $90^{\circ}$
in a gradual way, forming lines resembling circle segments while
maintaining an essentially constant separation.  Note that the two
innermost stream lines, as they leave the critical-state in the
concave corner, they cross over into the Meissner-state area.  This is
a clear manifestation of the non-locality of the governing equations,
and cannot be found within the bulk case of the Bean model.

At $H_a=0.6$ the domains of constant current density have grown in
size mainly by penetrating deeper, but also by filling larger parts of
all corners.  All the features commented for $H_a=0.3$ are here still
present.  In addition, the current now bends quite sharply at the
convex corners, where the d-lines become increasingly more well-defined.  

\begin{figure}[t]
  \centering
  \includegraphics[width=8.5cm]{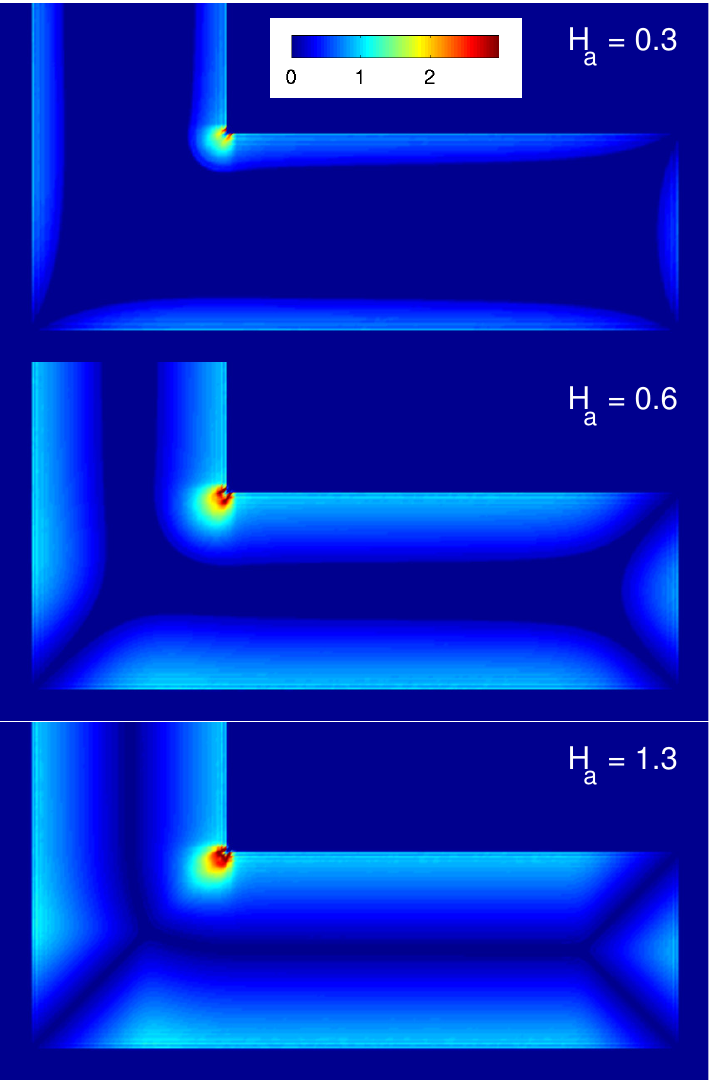} 
  \caption{
    \label{fig:E}
    Distribution of electric field $E$ at applied magnetic fields $H_a=0.3$, 0.6 and 1.3 
    during ramping of the magnetic field. The images are color coded to show the 
    variation in the magnitude of $E$, see the color bar.
    Note that $E$ is not calculated in the vacuum.
  }
\end{figure}

\begin{figure}[t]
  \centering
  \includegraphics[width=6cm]{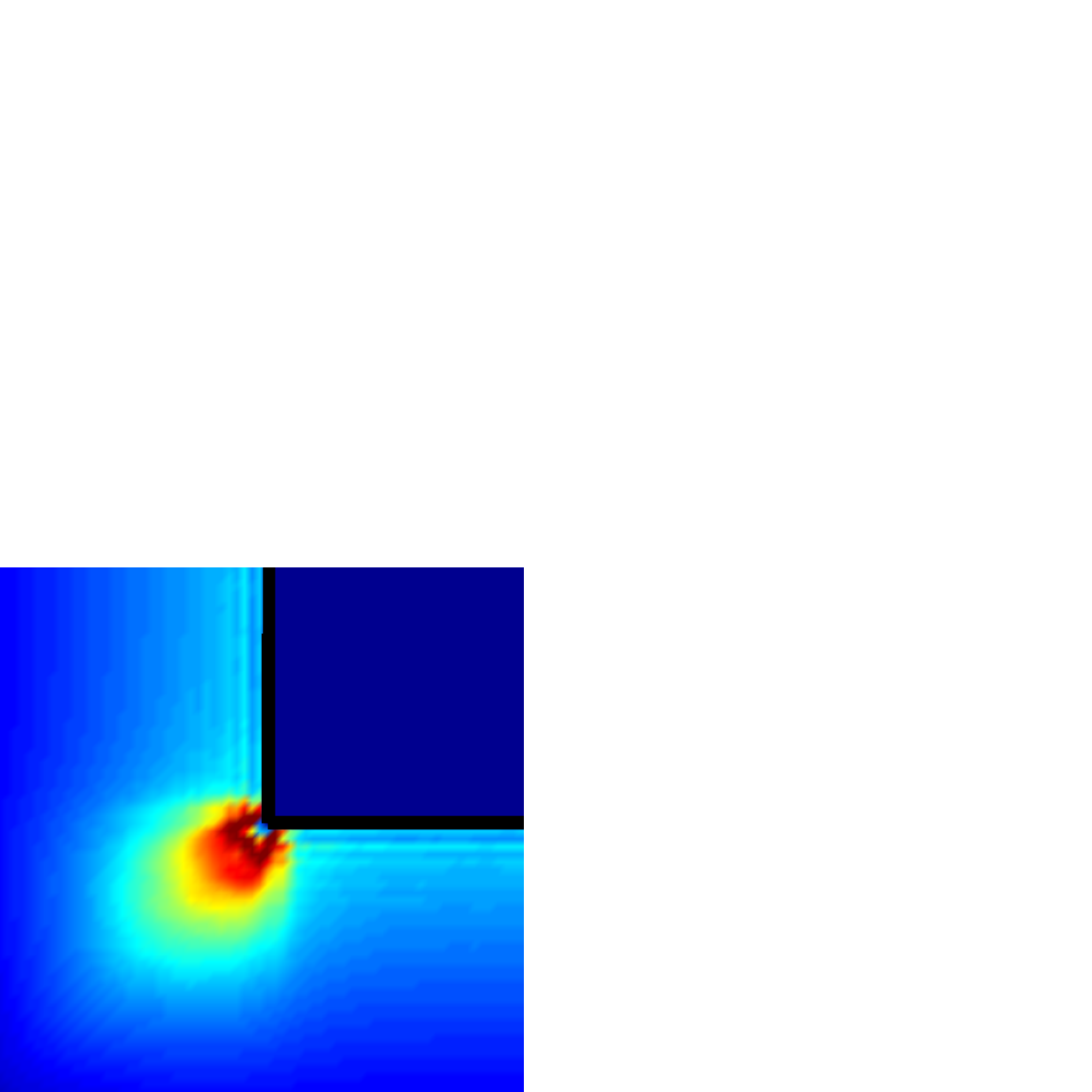} \  
  \includegraphics[width=0.6cm]{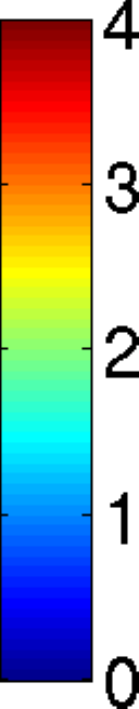} \\  
  \caption{
    \label{fig:E-part}
    The distribution of electric field $E$ close to the convex corner at $H_a = 1.3$.
  }
\end{figure}

At $H_a=1.3$, the full penetration state has been reached
and the current density is everywhere $J\approx 1$.
Note that in the central part of the sample, there are now several 
stream lines making small closed loops. 
At he same time, the d-line extending from the large convex corner is
longer than the others. This overall pattern is fully compatible with
the critical state model rule for constructing equidistant stream
lines. In particular, one finds that the length of the long d-line 
should be a factor $2(2-\sqrt{2})$, or 17\% longer than the others, in very good
agreement with the simulation result. Moreover, the construction
implies that the curved part of the central d-line consists of two
parabolic curves meeting at the end of the long d-line.

Figure~\ref{fig:E} shows the distribution of the electric field
magnitude $E$.  At the straight edges, the results are as expected
from the critical state model in an infinite strip, where the $E$-field
grows almost linearly from the flux front towards the
edges \cite{brandt95}.  In the Meissner state are $E=0$.
Also in agreement with the critical state model is
the strong suppression of $E$ close to the convex corners, where
at the d-lines $E=0$, signifying absence of any flux traffic. 
However,the most striking feature in the $E$-maps is the spot of very high
field at the concave corner, which is strongly present in all three
panels. Fig.~\ref{fig:E-part} focuses at this spot 
at $H_a=1.3$. The results appears similar to the numerical result of 
Ref.~\cite{schuster96}, having a maximum value $E\sim 4.5$, compared
to $E\sim 1$ at the straight edges.  This enhanced electrical field is
a sign of intensive flux traffic through the concave corner.

Since high electrical fields are known to trigger thermomagnetic
avalanches \cite{mints96,rakhmanov04,denisov05}, it is thus clear that
thin superconductors with concave corners are far more susceptible for
such dramatic events to occur than samples with only convex corners or
without corners at all.

As mentioned, the current stream lines near the concave corner appears
to be nearly circular.  Yet, the simulation results shown in
Figs.~\ref{fig:E} and \ref{fig:E-part} deviate from the $E$-field generated by perfectly
circular stream lines, as calculated in Ref.~\cite{schuster96}.  The
deviation manifests in three important ways.  First, the field value
at the edge is finite, and is not showing the divergent $E \sim 1/r$
behavior.  Secondly, the maximum $E$ is not found exactly at the
corner, but in two spots located on each side of the corner.  Thus,
the traffic of flux into the sample goes through a wider region than
just a singular point, explaining why $E$ is not divergent.  Thirdly,
the $E$-field is smooth everywhere inside the sample, and in
particular there is no discontinuity in $E$ at angles $0^\circ$ and
$90^\circ$.  Hence, the stream lines obtained from the simulations are
not simply circular segments connected with straight lines. This
motivates a closer inspection of the flow pattern.

\begin{figure}[t]
  \centering
  \includegraphics[width=7cm]{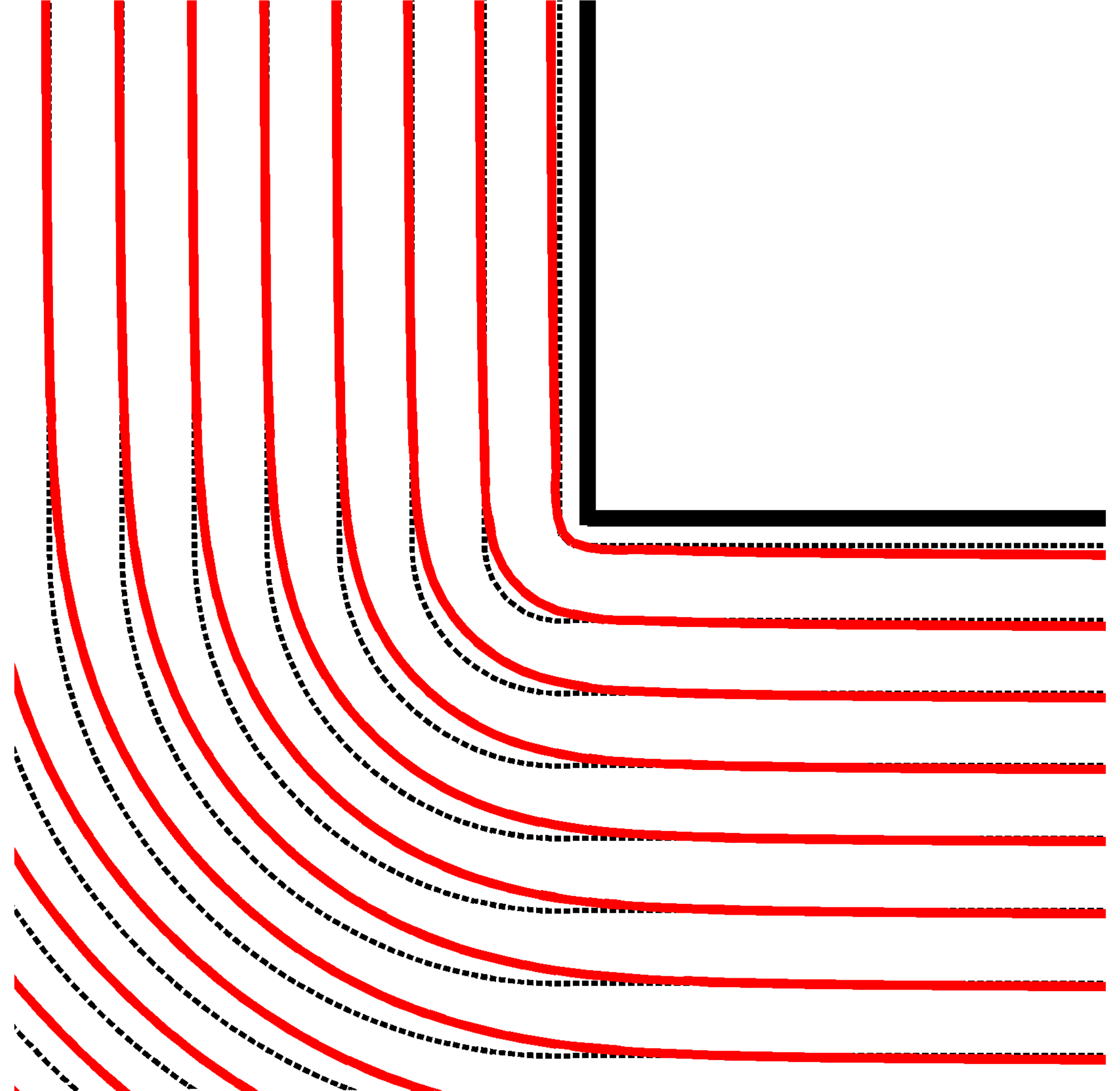} 
  \caption{
    \label{fig:sim-circ}
    The current stream lines resulting from the simulation (full red line)
    and for comparison, stream lines including circular segments (dotted black line).
  }
\end{figure}

Shown in Fig.~\ref{fig:sim-circ} is the detailed current stream line
pattern near the concave corner at the applied field $H_a=1.3$.  For
direct comparison the figure shows also stream lines shaped as
concentric circular segments connecting to straight lines running
parallel to the edges, and having a spacing corresponding to $J=1$.
The two patterns clearly deviate, as the calculated stream lines are
compressed as they pass the corner, implying that there is an enhanced
current density, $J>1$, in the region.  This is indeed consistent with
the electrical field map in Fig.~\ref{fig:E-part}, where the highest
electrical field is $E\sim 4.5$ giving a sheet current of
$4.5^{1/29}=1.05$, i.e., a 5\% increase in the current density.

Notice also that the calculated stream lines begin to curve some
distance before they meet the corner sector, thus the region of curves
lines cover a sector wider than $90^{\circ}$.
In this way,  the stream lines
change direction more gradually,  and the unphysical
discontinuity in $E$ mentioned earlier,  is avoided. 
We also note that the stream line pattern in Fig.~\ref{fig:sim-circ}
deviates from the analytically derived result presented in
Ref.~\cite{friesen01}, where the stream lines make a sharp bend, near
45$^\circ$, midway into the corner. Thus, it is evident that both the
nonlinearity and nonlocality of the governing equations are important
for the outcome of the modeling. One open question that remains is
how the results would change in the limit $n\to \infty$. The
numerical calculation of this work does not give a definite answer,
but since the applied creep exponent $n=29$ is rather high, we expect
that the results would be qualitatively the same.

It is appropriate now to compare the simulations with results of MOI
experiments.

\begin{figure}[t]
  \centering
  \includegraphics[width=8.5cm]{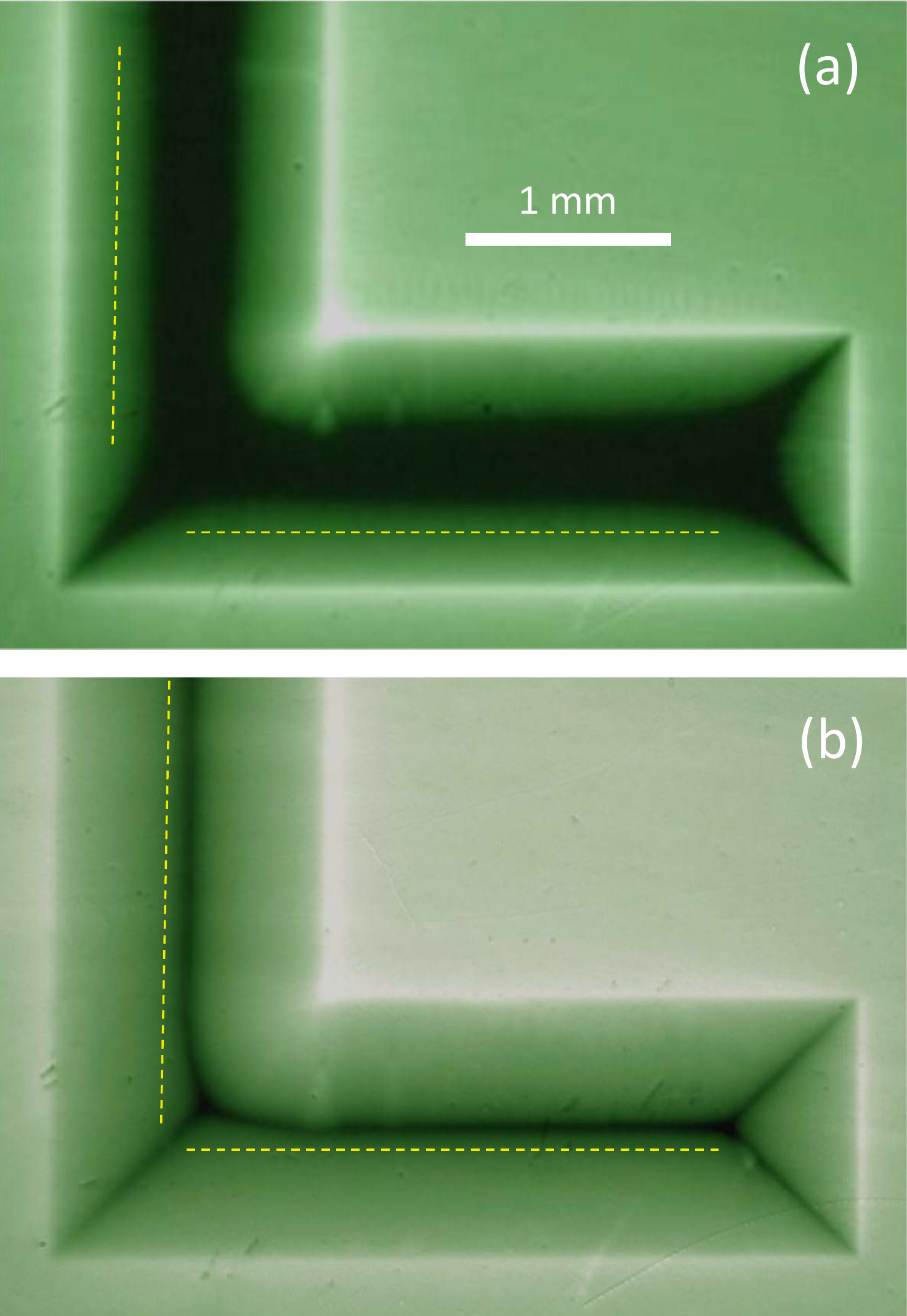} \
  \caption{
    \label{fig:MO}
    Magneto-optical images of flux penetration in a 
    YBa$_2$Cu$_3$O$_x$ film at 45~K. The images in (a) and (b) 
    were recorded at applied fields of
    $B_a$ = 16.5~mT and 44~mT, respectively.
  }
\end{figure}

\section{Experiments}

A film of YBa$_2$Cu$_3$O$_x$ was made by laser ablation on a (100)
SrTiO$_3$ substrate. Details of the preparation can be found in
Ref.~\cite{vase91}. The sample  has a
thickness of $d = 300$ nm with the c-axis oriented perpendicular to
the film plane. The critical temperature $T_c$, measured by magnetic
susceptibility, was 89.9 K. The critical current density of the film
is $j_c = 8.0 \times 10^{10}$ A/m$^2$ at 45~K, the temperature where
the magneto-optical images were recorded.

The MOI investigation was performed using a
bismuth substituted ferrite garnet film with in-plane magnetization as
Faraday rotating sensor \cite{helseth01}, placed directly on
the sample surface. The sample was mounted on the cold finger of a
continuous He flow cryostat with an optical window. Images of the flux
distribution were recorded with a digital camera through a polarized
light microscope using crossed polarizers.  In this way the image
brightness represents the magnitude of the flux density. For more details
of the method and setup, see Ref.~\cite{goa03}.

Shown in Fig.~\ref{fig:MO} are two images of the flux penetration in
the sample after an initial zero-field cooling to 45~K. Then the
applied perpendicular magnetic field was slowly increased, and the
images in (a) and (b) were recorded at $B_a$ of 16.5 and 44~mT,
respectively.  The image (a) was slightly contrast enhanced to allow
the location of the flux front to be clearly seen.

 In the concave corner region it is evident that the flux front has
advanced deeper than from the long straight edges of the sample.  The
“swollen” region covers a sector slightly exceeding $90^{\circ}$.  The
similarity between this experimental image and the simulated result
for $H_a =0.6$ shown in Fig.~\ref{fig:H} is striking.

Also on the other side of the sample, near the main convex corner,
 the experimental images show a non-trivial behavior. 
The flux front associated with the lower
edge  penetrates slightly deeper near the main convex corner.
The dashed line included in panel (b)  is parallel to the edge, 
and serve as a guide to the eye. The formation of the extra long
 d-line in the main corner is also clearly seen in the image (b).
The same “swelling” effect is seen to occur if one follows the
similar flux front down along the vertical part of the sample.
The agreement between these experimental results 
and the numerical simulations is striking down to the very fine details.

\section{Conclusions} 
In superconducting films, the electrodynamic non-locality implies that
calculations, in principle, must consider the whole sample and the
full magnetic history.  This is particularly evident at the concave
corner considered in this work, where the current stream line pattern
is nontrivial, unlike the convex corner where the stream line pattern
can be drawn using a simple Bean model procedure.  The flux creep
simulation conducted in this work shows that the stream line pattern
at the concave corner deviates from exact circles, and this small
deviation prevents unphysical jumps and infinities in the electrical
field. Nevertheless, the electrical field is very high at the concave
corner, and samples with such corners should thus be particularly
susceptible for nucleation of thermomagnetic avalanches.  The
magneto-optical experiment of this work shows striking similarity with
the simulation and verifies the necessity of properly including the
electrodynamic non-locality when modeling thin films in transverse
geometry.

\acknowledgments 
The financial support from the Research Council of Norway is 
gratefully acknowledged.\\

%

\end{document}